\input harvmac
\input amssym
\noblackbox

\mathchardef\varGamma="0100
\mathchardef\varDelta="0101
\mathchardef\varTheta="0102
\mathchardef\varLambda="0103
\mathchardef\varXi="0104
\mathchardef\varPi="0105
\mathchardef\varSigma="0106
\mathchardef\varUpsilon="0107
\mathchardef\varPhi="0108
\mathchardef\varPsi="0109
\mathchardef\varOmega="010A

\font\Scr=rsfs10

\def\scr#1{\hbox{\Scr #1}}

\def\Mt{{\kern1em\hbox{$\tilde{\kern-1em{\scr M}}$}}}

\font\sScr=rsfs7 
\def\sscr#1{\hbox{\sScr #1}}


\lref\ds{
M.~Dine and N.~Seiberg,
``Is The Superstring Weakly Coupled?,''
Phys.\ Lett.\  B162 (1985) 299.
}

\lref\heterotic{
L.~J.~Dixon and J.~A.~Harvey,
``String Theories In Ten-Dimensions Without Space-Time Supersymmetry,''
Nucl.\ Phys.\  B274 (1986) 93\semi
L.~Alvarez-Gaum\'e, P.~H.~Ginsparg, G.~W.~Moore and C.~Vafa,
``An ${\rm O}(16) \times {\rm O}(16)$ Heterotic String,''
Phys.\ Lett.\  B171 (1986) 155.
}

\lref\ob{
A.~Sagnotti,
``Some properties of open string theories,''
arXiv:hep-th/9509080,
``Surprises in Open-String Perturbation Theory', arXiv:hep-th/9702093\semi
C.~Angelantonj,
``Non-tachyonic open descendants of the 0B string theory,''
Phys.\ Lett.\  B444 (1998) 309
[arXiv:hep-th/9810214]\semi
R.~Blumenhagen, A.~Font and D.~Lust,
``Tachyon-free orientifolds of type 0B strings in various dimensions,''
Nucl.\ Phys.\  B558 (1999) 159
[arXiv:hep-th/9904069]\semi
K.~Forger,
``On non-tachyonic $\Bbb{Z}_N\times \Bbb{Z}_M$ orientifolds of type 0B string theory,''
Phys.\ Lett.\  B469 (1999) 113
[arXiv:hep-th/9909010].
}

\lref\sugimoto{
S.~Sugimoto,
``Anomaly cancellations in type I D9-$\overline{{\rm D}9}$ system and the USp(32)  string theory,''
Prog.\ Theor.\ Phys.\  102 (1999) 685
[arXiv:hep-th/9905159].
}

\lref\bsb{
I.~Antoniadis, E.~Dudas and A.~Sagnotti,
``Brane supersymmetry breaking,''
Phys.\ Lett.\  B464 (1999) 38
[arXiv:hep-th/9908023]\semi
C.~Angelantonj,
``Comments on open-string orbifolds with a non-vanishing $B_{ab}$,''
Nucl.\ Phys.\  B566 (2000) 126
[arXiv:hep-th/9908064]\semi
G.~Aldazabal and A.~M.~Uranga,
``Tachyon-free non-supersymmetric type IIB orientifolds via  brane-antibrane systems,''
JHEP 9910 (1999) 024
[arXiv:hep-th/9908072]\semi
C.~Angelantonj, I.~Antoniadis, G.~D'Appollonio, E.~Dudas and A.~Sagnotti,
``Type I vacua with brane supersymmetry breaking,''
Nucl.\ Phys.\  B572 (2000) 36
[arXiv:hep-th/9911081].
}

\lref\ads{
I.~Affleck, M.~Dine and N.~Seiberg,
``Calculable Nonperturbative Supersymmetry Breaking,''
Phys.\ Rev.\ Lett.\  52 (1984) 1677,
``Dynamical Supersymmetry Breaking In Supersymmetric QCD,''
Nucl.\ Phys.\ B241 (1984) 493.
}

\lref\iss{
K.~Intriligator, N.~Seiberg and D.~Shih,
``Dynamical SUSY breaking in meta-stable vacua,''
JHEP 0604 (2006) 021
[arXiv:hep-th/0602239].
}

\lref\ddgr{
S.~Dimopoulos, G.~R.~Dvali, R.~Rattazzi and G.~F.~Giudice,
``Dynamical soft terms with unbroken supersymmetry,''
Nucl.\ Phys.\ B510 (1998) 12
[arXiv:hep-ph/9705307]\semi
M.~A.~Luty,
``Simple gauge-mediated models with local minima,''
Phys.\ Lett.\  B414 (1997) 71
[arXiv:hep-ph/9706554].
}

\lref\issii{
S.~Franco and A.~M.~Uranga,
``Dynamical SUSY breaking at meta-stable minima from D-branes at obstructed geometries,''
JHEP 0606 (2006) 031
[arXiv:hep-th/0604136]\semi
H.~Ooguri and Y.~Ookouchi,
``Landscape of supersymmetry breaking vacua in geometrically realized gauge theories,''
Nucl.\ Phys.\  B755 (2006) 239 
[arXiv:hep-th/0606061],
``Meta-stable supersymmetry breaking vacua on intersecting branes,''
Phys.\ Lett.\  B641 (2006) 323
[arXiv:hep-th/0607183]\semi
V.~Braun, E.~I.~Buchbinder and B.~A.~Ovrut,
``Dynamical SUSY breaking in heterotic M-theory,''
Phys.\ Lett.\  B639 (2006) 566
[arXiv:hep-th/0606166],
``Towards realizing dynamical SUSY breaking in heterotic model building,''
JHEP 0610 (2006) 041
[arXiv:hep-th/0606241]\semi
S.~Ray,
``Some properties of meta-stable supersymmetry-breaking vacua in Wess-Zumino models,''
Phys.\ Lett.\  B642 (2006) 137
[arXiv:hep-th/0607172]\semi
S.~Franco, I.~Garcia-Etxebarria and A.~M.~Uranga,
``Non-supersymmetric meta-stable vacua from brane configurations,''
JHEP 0701 (2007) 085
[arXiv:hep-th/0607218]\semi
S.~Forste,
``Gauging flavour in meta-stable SUSY breaking models,''
Phys.\ Lett.\  B642 (2006) 142
[arXiv:hep-th/0608036]\semi
A.~Amariti, L.~Girardello and A.~Mariotti,
``Non-supersymmetric meta-stable vacua in ${\rm SU}(N)$ SQCD with adjoint matter,''
JHEP 0612 (2006) 058
[arXiv:hep-th/0608063]\semi
I.~Bena, E.~Gorbatov, S.~Hellerman, N.~Seiberg and D.~Shih,
``A note on (meta)stable brane configurations in MQCD,''
JHEP 0611 (2006) 088
[arXiv:hep-th/0608157]\semi
C.~Ahn,
``Brane configurations for nonsupersymmetric meta-stable vacua in SQCD with adjoint matter,''
Class.\ Quant.\ Grav.\  24 (2007) 1359
[arXiv:hep-th/0608160],
Phys.\ Lett.\  B647 (2007) 493
[arXiv:hep-th/0610025]\semi
M.~Eto, K.~Hashimoto and S.~Terashima,
``Solitons in supersymmety breaking meta-stable vacua,''
JHEP 0703 (2007) 061
[arXiv:hep-th/0610042]\semi
R.~Argurio, M.~Bertolini, S.~Franco and S.~Kachru,
``Gauge/gravity duality and meta-stable dynamical supersymmetry breaking,''
JHEP 0701 (2007) 083
[arXiv:hep-th/0610212],
``Metastable vacua and D-branes at the conifold,''
arXiv:hep-th/0703236\semi
M.~Aganagic, C.~Beem, J.~Seo and C.~Vafa,
``Geometrically induced metastability and holography,''
arXiv:hep-th/0610249\semi
R.~Tatar and B.~Wetenhall,
``Metastable vacua, geometrical engineering and MQCD transitions,''
JHEP 0702 (2007) 020
[arXiv:hep-th/0611303]\semi
A.~Giveon and D.~Kutasov,
``Gauge symmetry and supersymmetry breaking from intersecting branes,''
arXiv:hep-th/0703135\semi
I.~Garcia-Etxebarria, F.~Saad and A.~M.~Uranga,
``Supersymmetry breaking metastable vacua in runaway quiver gauge theories,''
arXiv:0704.0166 [hep-th].
}

\lref\dpp{
E.~Dudas, C.~Papineau and S.~Pokorski,
``Moduli stabilization and uplifting with dynamically generated F-terms,''
JHEP 0702 (2007) 028
[arXiv:hep-th/0610297]\semi
H.~Abe, T.~Higaki, T.~Kobayashi and Y.~Omura,
``Moduli stabilization, F-term uplifting and soft supersymmetry breaking terms,''
Phys.\ Rev.\  D75 (2007) 025019
[arXiv:hep-th/0611024]\semi
R.~Kallosh and A.~Linde,
``O'KKLT,''
JHEP 0702 (2007) 002
[arXiv:hep-th/0611183]\semi
O.~Lebedev, V.~Lowen, Y.~Mambrini, H.~P.~Nilles and M.~Ratz,
``Metastable vacua in flux compactifications and their phenomenology,''
JHEP 0702 (2007) 063
[arXiv:hep-ph/0612035].
}

\lref\fs{
W.~Fischler and L.~Susskind,
``Dilaton Tadpoles, String Condensates And Scale Invariance,''
Phys.\ Lett.\  B171 (1986) 383,
``Dilaton Tadpoles, String Condensates And Scale Invariance. 2,''
Phys.\ Lett.\  B173 (1986) 262.
}

\lref\dnps{
 E.~Dudas, G.~Pradisi, M.~Nicolosi and A.~Sagnotti,
``On tadpoles and vacuum redefinitions in string theory,''
Nucl.\ Phys.\  B708 (2005) 3
[arXiv:hep-th/0410101].
}

\lref\dm{
E.~Dudas and J.~Mourad,
``Brane solutions in strings with broken supersymmetry and dilaton tadpoles,''
Phys.\ Lett.\  B486 (2000) 172
[arXiv:hep-th/0004165]\semi
R.~Blumenhagen and A.~Font,
``Dilaton tadpoles, warped geometries and large extra dimensions for non-supersymmetric strings,''
Nucl.\ Phys.\  B599 (2001) 241 
[arXiv:hep-th/0011269].
}

\lref\discrete{
M.~Bianchi, G.~Pradisi and A.~Sagnotti,
``Toroidal compactification and symmetry breaking in open string theories,''
Nucl.\ Phys.\  B376 (1992) 365.
}

\lref\reviews{
E.~Dudas,
``Theory and phenomenology of type I strings and M-theory,''
Class.\ Quant.\ Grav.\  17 (2000) R41
[arXiv:hep-ph/0006190]\semi
C.~Angelantonj and A.~Sagnotti,
``Open strings,''
Phys.\ Rept.\  371 (2002) 1
[Erratum-ibid.\  376 (2003) 339]
[arXiv:hep-th/0204089].
}

\lref\Sdual{
S.~Elitzur, A.~Giveon, D.~Kutasov and D.~Tsabar,
``Branes, orientifolds and chiral gauge theories,''
Nucl.\ Phys.\  B524 (1998) 251
[arXiv:hep-th/9801020]\semi
E.~Witten,
``Baryons and branes in anti de Sitter space,''
JHEP 9807 (1998) 006
[arXiv:hep-th/9805112]\semi
A.~M.~Uranga,
``Comments on non-supersymmetric orientifolds at strong coupling,''
JHEP 0002 (2000) 041
[arXiv:hep-th/9912145]\semi
 A.~Hanany and B.~Kol,
``On orientifolds, discrete torsion, branes and M theory,''
JHEP 0006 (2000) 013
[arXiv:hep-th/0003025]\semi
Y.~Hyakutake, Y.~Imamura and S.~Sugimoto,
``Orientifold planes, type I Wilson lines and non-BPS D-branes,''
JHEP 0008 (2000) 043
[arXiv:hep-th/0007012].
}

\lref\DudasNV{
E.~Dudas and J.~Mourad,
``Consistent gravitino couplings in non-supersymmetric strings,''
Phys.\ Lett.\  B514 (2001) 173
[arXiv:hep-th/0012071]\semi
G.~Pradisi and F.~Riccioni,
``Geometric couplings and brane supersymmetry breaking,''
Nucl.\ Phys.\  B615 (2001) 33
[arXiv:hep-th/0107090]\semi
J.~H.~Schwarz and E.~Witten,
``Anomaly analysis of brane-antibrane systems,''
JHEP 0103 (2001) 032
[arXiv:hep-th/0103099].
}

\lref\GinspargWR{
P.~H.~Ginsparg and C.~Vafa,
``Toroidal Compactification of Nonsupersymmetric Heterotic Strings,''
Nucl.\ Phys.\  B289 (1987) 414\semi
V.~P.~Nair, A.~D.~Shapere, A.~Strominger and F.~Wilczek,
``Compactification of the Twisted Heterotic String,''
Nucl.\ Phys.\  B287 (1987) 402.
}

\lref\Sen{
A.~Sen,
``Non-BPS states and branes in string theory,''
arXiv:hep-th/9904207.
}

\lref\hw{
P.~Horava and E.~Witten,
``Heterotic and type I string dynamics from eleven dimensions,''
Nucl.\ Phys.\  B460 (1996) 506 
[arXiv:hep-th/9510209].
}

\lref\adds{
I.~Antoniadis, G.~D'Appollonio, E.~Dudas and A.~Sagnotti,
``Partial breaking of supersymmetry, open strings and M-theory,''
Nucl.\ Phys.\  B553 (1999) 133
[arXiv:hep-th/9812118].
}


\Title{\vbox{
\rightline{CPHT-RR 017.0417}
\rightline{DFTT 2007/5}
\rightline{LPT--Orsay 07/23}
}}
{\vbox{
\centerline{Metastable String Vacua}
}}

\centerline{Carlo Angelantonj$^\dagger$ and Emilian Dudas${}^\ddagger{}^\star$}

\medskip
\centerline{\it $^\dagger$ Dipartimento di Fisica Teorica and INFN Sezione di Torino}
\centerline{\it Via P. Giuria 1, I--10125 Torino}
\centerline{\it $^\ddagger$ CPhT Ecole Polytechnique, CNRS, F--91128 Palaiseau Cedex}
\centerline{\it $^\star$ Laboratoire de Physique Th\'eorique, Universit\'e Paris-Sud, F--91405 Orsay}

\vskip 0.6in

\centerline{\bf Abstract}

\noindent
We argue that tachyon-free type I string vacua with
supersymmetry breaking in the open sector at the string scale can be
interpreted, via S and T-duality arguments, as metastable vacua of the
supersymmetric type I superstring. The dynamics of the process can
be partly captured via nucleation of brane-antibrane
pairs out of the non-supersymmetric vacuum and subsequent tachyon
condensation.

\Date{April 2007} 


\newsec{Introduction and conclusions}

It is a widespread belief that all perturbative string constructions with broken supersymmetry are unstable and that the dynamics universally drives them towards trivial configurations 
\ds. Typically, the simplest sign of instability of non-supersymmetric string vacua is the presence of tachyonic excitations, at least in some regions of moduli space. Although, in the past tachyon-free ten-dimensional vacua with broken supersymmetry have been proposed 
\refs{\heterotic,\ob,\sugimoto,\bsb} possibly violating the standard lore, it was soon evident that most of these string vacua develop tachyonic instabilities once some dimensions are compactified. For instance, the ${\rm O}(16)\times {\rm O}(16)$ heterotic model  \heterotic\ is continuously related to its tachyonic cousins after proper Wilson lines are introduced in nine dimensions \GinspargWR, while for the circle reduction of the $0'{\rm B}$ model \ob\ either the winding or the momentum excitations of the closed-string tachyon are still present after the orientifold projection, and actually become tachyonic in the small or large radius region of moduli space, respectively. 

The so-called type I vacua with brane supersymmetry breaking \refs{\sugimoto,\bsb}, however, seem to be non-tachyonic, and thus stable, in any space-time dimension and in any corner of moduli space, thus offering a notable counter example to this common belief \ds. These models are characterised by a supersymmetric closed-string sector, while supersymmetry is explicitly broken in the open-string sector at the string scale, where bosonic and fermionic excitations are assigned different representations with of the Chan-Paton gauge group. Although the presence of gauge singlet fermions hints to the fact that the vacuum is already in its broken phase, where supersymmetry is non-linearly realised \DudasNV, there is no obvious candidate for a supersymmetric vacuum configuration to which it could decay into.

Whether or not these models are quantum mechanically stable is an open issue that we shall try to elucidate in the present letter. Actually, the construction of metastable vacua in field theories with rigid supersymmetry \iss\ has acquired some interest, and it is believed that they are more natural than traditional models with dynamical supersymmetry breaking \ads\
(see \ddgr\ for earlier constructions of metastable vacua). Some proposals to extend the field theory constructions in \iss\ to string theory using D-branes at orbifold singularities have been suggested \issii , while in \dpp\ it was argued that metastable vacua could play an active role in attempts to stabilise moduli. Despite much progress in the field theory and/or string theory constructions with metastable phases, identifying a full-fledged string theory vacuum of this type is still an important unsolved problem. Clearly, around such a metastable vacuum the non-supersymmetric spectrum should be free of tachyonic excitations, precisely as in the case for orientifolds with brane supersymmetry breaking \refs{\sugimoto,\bsb}. It is then natural to propose that these vacua actually represent metastable local minima in the moduli space, where the true global minimum would correspond to the supersymmetric type I superstring. 
The purpose of this note is to collect some evidence in favour of this conjecture. In fact, we shall show that the  models in \refs{\sugimoto,\bsb} are naturally driven towards strong coupling. A Montonen-Olive duality then leads to a natural perturbative description in terms of type I superstring with pairs of branes and anti-branes that are expected to decay to the SO(32) superstring after brane and anti-brane annihilation. We shall also show how this dynamics could be partly captured by the condenstation of tachyons on the pairs of branes and anti-branes.


\newsec{Non-BPS string vacua and strong coupling}

Orientifold models are the subject of an intense activity, since their perturbative definition offers interesting new possibilities for low-energy phenomenology. These models have a very interesting geometrical description in terms of D-branes and orientifold planes, extended objects that carry a charge with respect to appropriate R-R potentials and have a tension proportional to the charge itself. Typically, tensions and charges of D-brane and O-planes saturate a BPS bound, so that individually they preserve a certain half of the original supersymmetries of the closed-string theory, depending on the relative sign of their tension and charge. For D-branes tension and charge are both positive, while two types of O-planes can be present in perturbative string vacua: those with negative tension and charge, here denoted ${\rm O}p_-$-planes, and those with positive tension and positive charge, here denoted ${\rm O}p_+$-planes\foot{Notice that we have here changed our original conventions  \reviews\ to those widely used in the current literature.}. In addition, there are of course anti-D-branes and anti-O-planes, with identical tension and opposite R-R charges. Moreover, using non-perturbative string dualities, a rich zoo of similar extended objects emerges \Sdual\ that will be used in the following sections to support our conjecture. 

The consistency of orientifold constructions and a number of their most amusing features may be traced to the relation to suitable parent models of oriented closed strings, from which their spectra can be derived \reviews. In this procedure, a special role is played by tadpole conditions for R-R and NS-NS states. Although space-time supersymmetry relates the two tadpole conditions, they are completely different in nature. In fact, while the former are to be regarded as global neutrality conditions for R-R charges, and are usually linked to gauge and gravitational anomalies, the latter simply force the configuration of D-branes and O-planes to be globally massless. As a result, while the R-R tadpoles have always to be cancelled in a consistent vacuum configuration, in principle NS-NS ones can be relaxed, thus calling for a background redefinition \refs{\fs,\dnps} whose proper implementation in string theory, however, is not fully understood. 

This difference between R-R and NS-NS tadpoles turns out to play an important role in a class of models with broken supersymmetry. In these constructions \refs{\sugimoto,\bsb}, the closed-string sector is classically supersymmetric, whereas supersymmetry is broken at the string scale on some stack of D-branes. Geometrically, these models always involve 
${\rm O}p_+$ planes together with an appropriate number of anti-branes, termed $\overline{{\rm D}p}$-branes in the following, whose negative R-R charge compensates that of the ${\rm O}p_+$ planes. In the simplest known example \sugimoto , the ten-dimensional closed-string sector encoded in the torus and Klein-bottle partition functions\foot{We are omitting in all vacuum amplitude an overall normalisation factor that however does not affect our qualitative description.}
\eqn\closed{
{\scr T} = {\textstyle{1\over 2}} \int_{\sscr F} {d^2\tau \over \tau_2^6} \, {| V_8 - S_8 |^2 \over |\eta |^{16}}\,, \qquad 
{\scr K} = {\textstyle{1\over 2}} \int_0^\infty {d\tau_2 \over \tau_2^6}\, {V_8 - S_8 \over \eta^8}\,,
}
is as in the supersymmetric type I superstring, while in the open-string sector encoded in the annulus and M\"obius-strip amplitudes
\eqn\open{
{\scr A} = {\textstyle{1\over 2}} N^2 \int_0^\infty {d t \over t^6} \, {V_8 - S_8 \over \eta^8}\,, \qquad 
{\scr M} = {\textstyle{1\over 2}} N \int_0^\infty {dt \over t^6} \, {\hat V_8 + \hat S_8 \over \hat\eta ^8}\,,
}
a crucial sign difference in front the of NS sector in ${\scr M}$ yields a D-brane spectrum with broken supersymmetry. In fact, the orientifold projection is in this case $\varOmega' = - \varOmega (-1)^{F}$, where $(-1)^F$ is the space-time fermion number, so that  the massless gauge bosons have symmetric Chan-Paton matrices,  $\lambda_b = -\gamma_{\varOmega} \lambda^T_b \gamma_{\varOmega}^{-1} =  \lambda^T_b$, while the space-time fermions have anti-symmetric Chan-Paton matrices, $\lambda_f = +\gamma_{\Omega} \lambda^T_f \gamma_{\Omega}^{-1} =  - \lambda^T_f$. As a result, after setting $N=32$ as required by the cancellation of the R-R tadpole, the open-string spectrum has gauge group USp(32) and fermions in the reducible ${\bf 496}={\bf 495}+{\bf 1}$ anti-symmetric representation, consistently with the cancellation of ten-dimensional gauge and gravitational irreducible anomalies.

As usual, the transverse-channel M\"obius-strip amplitude
\eqn\trasv{
\Mt = N \int_0^\infty d\ell\, {\hat V_8 + \hat S_8 \over \hat\eta^8}
}
clearly spells out the nature of O-planes and D-branes involved in the construction that, as anticipated, are ${\rm O}9_+$-planes and $\overline{{\rm D}9}$-branes. This non-BPS configuration breaks explicitly all supersymmetries directly at the string scale, and seems not continuously connected to any supersymmetric vacuum. Notice that no tachyonic excitations are present in the open-string sector, thus suggesting that this vacuum configuration is locally, classically, stable. The quantum dynamics of this and related systems is, to the best of our knowledge, still an open question.

The impossibility of cancelling the NS-NS tadpole in these non-BPS configurations induces a tree-level potential in the low-energy effective action 
\eqn\dilatontadpole{
{\scr V} \sim {N+32 \over (\alpha ')^5} \, e^{-\phi}\,. 
}
While crucial in order to couple consistently a non-supersymmetric open-string spectrum to a supersymmetric bulk\foot{On the branes supersymmetry is actually realised non-linearly \DudasNV\ and the dilaton tadpole is the leading term in the expansion of the Volkov-Akulov action for the goldstino, the gauge-singlet spinor present among the open-string excitations.}, this potential is incompatible with a maximally symmetric Minkowski space-time, and in fact leads to a ``spontaneous compactification'' to nine dimensions, with a manifest SO(1,8) Poincar\'e symmetry. More specifically, the metric and the dilaton field read \dm
\eqn\background{
\eqalign{
e^{\phi} & = \ e^{\phi_0} |u|^{2/3} e^{3 u^2 /4} \ ,
\cr
ds^2  &= \  |u|^{4/9} e^{\phi_0/2} e ^{u^2
  /4} \eta_{\mu \nu} dx^{\mu}  dx^{\nu} +|u|^{-2/3} e^{-\phi_0} e ^{- 3 u^2 /4} dx^2 \ ,
\cr} 
}
in the string frame, where $u$ is the ``internal'' coordinate. Notice that in the Einstein frame the dilaton tadpole is proportional to $ e^{3 \phi /2}$, and hence one would naively expect the theory to be driven towards zero string coupling, with $g_{\rm s} = e^\phi$.
Actually, this is not the case, and inspection of the solution \background\ shows that
this vacuum configuration necessarily enters a strong coupling regime for large $u$. 
This clearly suggests that the perturbative description is at best incomplete. Another hint pointing towards the inevitable presence of a strongly coupled phase comes from the analysis of the gauge theory on the D-branes. After a suitable reduction to four dimensions, the light excitations comprise gauge bosons and six scalars in the adjoint of USp(32) together with four Weyl fermions in the 496-dimensional anti-symmetric representation. This gauge theory is clearly asymptotically free, and its coupling becomes strong at low energies. 
To summarise, these non-BPS orientifolds are naturally driven towards a phase of 
strong coupling, and, as we shall see in the following sections, our conjecture is that non-perturbatively these vacua are metastable states of the supersymmetric type I superstring.

\vskip .3 truein
\vbox{
\settabs = 6 \columns
\+ &  \vbox{\hrule height1pt width250pt} \cr
\+ & & $(\theta_{\rm NS}\,,\, \theta_{\rm R})$ & R-R charge & $G_{\rm CP}$ \cr
\+ &  \vbox{\hrule width250pt} \cr
\+ & ${\rm O}p_-$ & $(0,0)$ & $-2^{p-5}$ & ${\rm SO}(2n)$ \cr
\+ & ${\rm O}p_+$ & $({1\over 2},0)$ & $+2^{p-5}$ & ${\rm USp}(2n)$ \cr
\+ & $\widetilde{{\rm O}p}_-$ & $(0,{1\over 2})$ & ${1\over 2}-2^{p-5}$ & ${\rm SO}(2n+1)$ \cr
\+ & $\widetilde{{\rm O}p}_+$ & $({1\over 2},{1\over 2})$ & $+2^{p-5}$ &${\rm USp}(2n)$ \cr
\+ & \vbox{\hrule height1pt width250pt} \cr
\cleartabs
\centerline{\ninepoint{\bf Table 1.} The four types of O-planes for $p\le 5$.}
}
\vskip .2truein


\newsec{S-duality, supersymmetry breaking and metastable states}

In the previous section we have introduced two different types of O-planes that exist in perturbative string theory. We have called them ${\rm O}p_{\pm}$ planes where the suffix refers to the sign of their tension and charge. Actually, the difference between these two types of orientifold planes resides in a discrete $B_{ab}$ background, always allowed by the orientifold projection \discrete, that implies the possibility of having a non-trivial discrete holonomy for the NS-NS $B$ field
\eqn\holonomyB{
\theta_{\rm NS} =\int_{\Bbb{RP}^2} {B_2\over 2\pi} =  {\textstyle{1\over 2}}\,.
}
The holonomy contributes to a term $e^{2 i \pi \theta_{\rm NS}}$ to the $\Bbb{RP}^2$ amplitude and thus introduces and additional minus sign responsible for the  exchange of ${\rm O}p_{+}$ and ${\rm O}p_{-}$ planes. Actually, it was realised that also R-R field could have a non-trivial discrete holonomy, that would in turn yield new variants of orientifold planes. For instance, in the case of O3 planes one could allow for the holonomy
\eqn\holonomyRR{
\theta_{\rm R} = \int_{\Bbb{RP}^2} {C_2 \over 2\pi} = {\textstyle{1\over2}} \,.
}
As a result, there are four different types of orientifold planes characterised by the values of the holonomies $(\theta_{\rm NS} \,,\, \theta_{\rm R})$ and yield different types of gauge theories on stacks of D3-branes coincident with them, as summarised in table 1 \Sdual.

Unlike the ${\rm O}p_\pm$ cases, however, O-planes carrying a non-vanishing $\theta_{\rm R}$ holonomy cannot be described in perturbation theory since they involve a non-trivial R-R background. In fact, the ${\rm SL} (2,\Bbb{Z})$ duality of the type IIB superstring exchanges $\theta_{\rm NS}$ and $\theta_{\rm R}$, so that ${\rm O}3_-$ and $\widetilde{{\rm O}3}_+$ planes are fixed, while ${\rm O}3_+$ and $\widetilde{{\rm O}3}_-$ planes are interchanged. If we include D3 branes, the S-duality of type IIB becomes the Montonen-Olive duality for the ${\scr N}=4$ supersymmetric gauge theory living on their world-volume \Sdual. Notice, that an $\widetilde{{\rm O}3}_-$ plane has the same charge and tension as an ${\rm O}3_-$ plane with a stuck D3 brane on it, and indeed it was argued in \Sdual\ that in the strong coupling limit the ${\rm O}3_+$ plane with positive tension and positive charge is naturally described in terms of an ${\rm O}3_-$ together with a stuck D3. These are all the ingredients we need to describe the strong-coupling dynamics of the orientifold vacua introduced in the previous section. 

For simplicity, let us consider a local configuration of an ${\rm O}3_+$ plane with a number $m$ of $\overline{{\rm D}3}$ branes on it --- together with their images under $\varOmega$. Clearly this configuration is not BPS, and indeed the gauge theory on the anti-branes has gauge bosons and six scalars in the adjoint representation of a ${\rm USp} (2m)$ group while the four Weyl fermions are in the anti-symmetric representation. This is a local version of the model described in the the previous section and introduced in \refs{\sugimoto,\bsb}.
Although, strictly speaking, Montonen-Olive duality does not apply to this configuration, if the $\overline{{\rm D}3}$ branes are moved a distance $\delta \gg \sqrt{\alpha '}$ from the O-plane then supersymmetry is only mildly broken, and one can assume that S-duality is almost exact. Hence, the configuration of ${\rm O}3_+$ and $\overline{{\rm D}3}$ branes that is stable at weak coupling is naturally driven towards a strongly coupled regime where it is more conveniently described in terms of\foot{Notice that the $\overline{{\rm D}3}$ branes in the bulk have an ${\scr N}=4$ supersymmetric massless spectrum with gauge group ${\rm U} (m)$ that is self-dual.}  $\widetilde{{\rm O}3}_-$ and $\overline{{\rm D}3}$ or, better, in terms of a negatively charged ${\rm O}3_-$ plane plus $m$ physical  $\overline{{\rm D}3}$ and a stuck D3 brane. 

The vacuum energy of the initial configuration receives contribution entirely from the  M\"obius-strip  amplitude
\eqn\vacenw{
\eqalign{
\varLambda_{\rm weak} = - {\scr M}_{\rm weak} &=  - m \int_0^{\infty} {dt \over t^3} {{\hat V}_8 + {\hat S}_8 \over {\hat \eta}^8} \, \ e^{- 4 \pi t \delta^2 / \alpha '}  
\cr
&= - {m\over 4} \int_0^\infty {d \ell \over \ell^3} \, {\hat \theta_2^4 \over \hat\eta^{12}} \, e^{-2 \pi \delta^2/\alpha' \ell}
\cr 
& \sim - {m \, (\alpha ')^2 \over \pi^2 \delta^4} \, ,
\cr}
}
where the leading contribution originates from the exchange of massless closed-string states in the tree-level channel, and in going from the first to the second line we have used the standard relations between the proper times $t$ for the open-string propagation and $\ell$ for the closed-string propagation \reviews.

In the weakly coupled S-dual configuration, however, the $\overline{{\rm D}3}$ branes not only interact with the orientifold plane, but also with the stuck D3 brane, so that now both the annulus and M\"obius-strip diagrams contribute to the vacuum energy
\eqn\vacens{
\eqalign{
\varLambda_{\rm strong} &= - {\scr A}_{\rm strong} - {\scr M}_{\rm strong} 
\cr
&=
- 2 m \int_0^{\infty} {dt \over t^3} {O_8-C_8 \over \eta^8} \, e^{- \pi t \delta^2 / \alpha '} + m \int_0^{\infty} {dt \over t^3} {{\hat V}_8 +
{\hat S}_8 \over {\hat \eta}^8} \, \ e^{- 4 \pi t \delta^2 / \alpha '}  
\cr
&= -{m\over 2} \int_0^\infty {d\ell \over \ell^3} \, {\theta_2^4 \over \eta^{12}} \, e^{-2\pi \delta^2/\alpha '\ell} + {m\over 4} \int_0^\infty {d\ell \over \ell^3} \, {\hat\theta_2^4 \over \hat\eta^{12}} \, e^{-2 \pi \delta^2/\alpha ' \ell}
\cr
&\sim - {m (\alpha ')^2 \over \pi^2 \delta^4} \,.
\cr}
}
This configuration is clearly unstable since the $\overline{{\rm D}3}$ branes are attracted by the ${\rm O}3_-$ plane and the stuck D3 brane. However, in contrast with the original non-BPS configuration, for $\delta < \sqrt{\alpha '}$ a tachyonic mode now appears in the open-string spectrum and the $\overline{{\rm D}3}$'s and the stuck D3 tend to partially annihilate. 
In the next section we shall see how this local construction can be extended to vacuum configurations with brane supersymmetry breaking.

In the original non-BPS configuration the vacuum energy in the Einstein frame has a qualitative dependence on the string coupling constant of the form ${\scr V} \sim T - g_{\rm s} /\delta^4$ that indeed drives the system towards a non-perturbative regime. However, as $g_{\rm s}$ becomes strong, the non-BPS configuration has a natural weakly coupled description in terms of type I with pairs of branes and antibranes that is still characterised by a vacuum energy of the form 
${\scr V} \sim T - g_{\rm s} ' /\delta^4$. However, $g_{\rm s} ' = g_{\rm s}^{-1}$ is now very small and hence the corresponding vacuum energy is bigger, thus interposing an energy barrier between the original non-BPS configuration and the final type I superstring state. We are therefore led to conclude that the original non-BPS configuration, with ${\rm O}3_{+}$ plane and $\overline{{\rm D}3}$ branes,  is a locally metastable vacuum of a type IIB orientifold with ${\rm O}3_{-}$ planes. Clearly, this argument is somewhat qualitative, and
more detailed studies are needed in order prove that this non-BPS configuration is metastable.


\newsec{Strong coupling limit of vacua with brane supersymmetry breaking}

We can now use the strong coupling properties of the local model studied in the previous section to describe the dynamics of the non-BPS vacuum configuration of interest 
\refs{\sugimoto, \bsb}. In fact, let us consider the four-dimensional orientifold obtained by projecting the $T^6$ reduction of type IIB superstring by $\varOmega' = \varOmega I_6 (-1)^{F_{\rm L}}$, where $\varOmega$ is the standard orientifold projection, $I_6$ reverts the coordinates of the internal six-torus, and $(-1)^{F_{\rm L}}$ is the left-handed space-time fermion index. This orientifold introduces 64 ${\rm O}3_+$ planes at the 64 fixed points of the $\varOmega '$ orientifold together with 32 $\overline{{\rm D}3}$ branes needed to cancel the R-R tadpole.

The presence of a non-vanishing dilaton tadpole or, in turn, an attractive force between the 
${\rm O}3_+$ planes and the antibranes makes the configuration unstable and drives the model towards a strong coupling regime. If the $\overline{{\rm D}3}$ are placed in the bulk at a suitable distance from the O-planes, it is reasonable to assume that type IIB S-duality still holds, so that a weakly coupled description is in terms of 64 $\widetilde{{\rm O}3}_-$ planes, or in terms of 64 ${\rm O}3_-$ planes with 64 stuck D3 branes. This configuration is indeed allowed since the six Wilson lines 
\eqn\wilson{
\eqalign{
W_1 &= (1^{32},-1^{32}) \,, \qquad W_2 = (1^{16},-1^{16},-1^{16},1^{16}) \,,
\cr
W_3 &= (1^{8},-1^{8},-1^{8},1^{8}, -1^{8},1^{8},1^{8},-1^{8}) \,, \qquad \ldots
\cr} 
}
needed to distribute the D3 branes on the orientifold planes, have positive determinant and mutually commute when acting on spinors. One can then decompactify this configuration and at the same time undo these Wilson lines, so that the D3 branes can be brought together to yield an ${\rm SO} (64)$ gauge group. 

Finally, the 32 pairs of branes and antibranes annihilate via open-string tachyon condensation \Sen\ and one is left with the type I superstring with negatively charges
O-planes and 32 D-branes with gauge group SO(32).

This strongly coupled dynamics of the USp(32) model and its connection with the type I superstring can be nicely captured to a large extent by tachyon condensation already in ten dimensions. Let us consider, in fact, the type I superstring with additional pairs of branes and antibranes. In the presence of the ${\rm O}9_-$ plane these have two possible ways to decay. Either they fully annihilate in pairs, or a pair of stuck D9-$\overline{{\rm D}9}$ branes is left with an O(1) gauge group on each world-volume.  Taking into account also the $N=32$ D9 branes of type I, one is altogether left with $p=1$ stuck antibranes and $33=N+q$ branes whose one-loop amplitudes read
\eqn\acondens{
{\scr A} = \int_0^{\infty} {dt \over t^6} \, {1 \over \eta^8} \,
  \left[  {\textstyle{1 \over 2}} \left( (N+q)^2  + p^2 \right) (V_8 - S_8) + (N+q) \, p \, (O_8 - C_8) \right] \,, 
}
and
\eqn\mcondens{
{\scr M}  =  {\textstyle{1\over 2}} \int_0^{\infty} {dt \over t^6} \,{1 \over {\hat \eta}^8}\, \left[ - (N+q) \, ({\hat V}_8 - {\hat S}_8)  -  p\,  ({\hat V}_8 + {\hat S}_8) \right] \,. 
}
The light spectrum now comprises ${1\over 2} \, 33\cdot 32=528$ gauge bosons on the D9 branes, $32+1$ tachyons, denoted $T_{32}$ and $T_1$, $496 + 32 +1$ left-handed Majorana-Weyl fermions, denoted $\psi^{\rm L}_{496}$, $\psi^{\rm L}_{32}$ and $\psi^{\rm L}_{1}$, and $32+1$ right-handed Majorana-Weyl fermions, denoted $\lambda^{\rm R}_{32}$ and $\lambda^{\rm R}_{1}$. These massless excitations are compatible both with a SO(33) and a USp(32) gauge group. From the point of view of the former, tachyon condensation breaks it to its SO(32) subgroup and theory becomes the supersymmetric type I. However, we can interpret the end-point of tachyon condensation also from the viewpoint of USp(32) gauge group. In this case,  condensing the singlet tachyon, $\langle T_{1} \rangle \not= 0$, yields mass terms for the 33 non-chiral fermions and for the 32 $(N,p)$ tachyons $T_{32}$ through couplings of the form $T_1 \, \psi^{\rm L}_{32} \, \lambda^{\rm R}_{32}$, $T_1 \, \psi^{\rm L}_{1} \, \lambda^{\rm R}_{1}$ and $T_1^2 \, T_{32}^2$. 

As a result, the surviving massless modes are the 496 left-handed fermions $\psi^{\rm L}_{496}$ and 528 gauge bosons, precisely the massless content of the non-supersymmetric USp(32) gauge theory with chiral fermions in the anti-symmetric representation!


\newsec{S-duality in freely acting orbifolds with brane supersymmetry breaking}

Other non-BPS configurations similar to that discussed in section 2 have been proposed in the literature \bsb, and their fate is also an open question. Clearly, it would be nice if similar arguments based on S-duality could be applied also to these cases. Unfortunately, in most of the other models the non-supersymmetric branes are embedded in an ${\scr N}=2$ or ${\scr N}=1$ closed-string setting, and S duality is not  fully under control. For this reason, we shall study here a new vacuum partially related to that in \bsb, but where the various ingredients --- O-planes and D-branes --- are fairly separated in the transverse directions, and therefore do not interact strongly. The model is based on a freely acting $(T^4\times S^1\times S^1)/\Bbb{Z}_2$ orbifold of the type IIB superstring, where the single  $\Bbb{Z}_2$ generator $g$ reverts the sign of the four coordinates of the $T^4$
\eqn\reverts{
g:\ (X_6 \,,\, X_7 \,,\, X_8  \,,\, X_9 ) \to - (X_6 \,,\, X_7 \,,\, X_8  \,,\, X_9 )\,,
}
and simultaneously shifts the first $S^1$ coordinate
\eqn\shift{
g:  X_5 \to X_5 + \pi R \,,
}
by half of the length of the circle, while leaving untouched the $X_4$ coordinate of the second $S^1$. At the level of the type IIB superstring
\eqn\freelytorus{
\eqalign{
{\scr T} =& {\textstyle{1\over 2}} \int_{\sscr F} {d^2 \tau \over \tau_2^3} \, {1\over |\eta|^8}\, \left[ |Q_o + Q_v|^2 \varGamma^{(4,4)} \, \varGamma_{m,n} + |Q_o - Q_v|^2 \left|{2 \eta \over \theta_2}\right|^4\, (-1)^m \varGamma_{m,n} \right.
\cr
& \left. + 16  \, |Q_s + Q_c|^2 \, \left| {\eta\over \theta_4}\right|^4 \, \varGamma_{m,n+{1\over 2}} +16 \, |Q_s - Q_c|^2 \, \left| {\eta\over \theta_3}\right|^4 \, (-1)^m \varGamma_{m,n+{1\over 2}}
\right] \, \varGamma^{(1,1)}\,,
\cr}
}
it interpolates between ${\scr N}=2$ vacua and ${\scr N}=4$ vacua in the limit $R\to \infty$. Here we have used our standard notation \reviews\ for the $\Bbb{Z}_2$ characters
\eqn\char{
\eqalign{
Q_o &= V_4 O_4 - C_4 C_4\,,
\cr
Q_v &= O_4 V_4 - S_4 S_4\,,
\cr}
\qquad 
\eqalign{
Q_s &= O_4 C_4 - S_4 O_4\,,
\cr
Q_c &= V_4 S_4 - C_4 V_4\,,
\cr}
}
written in terms of SO(4) ones, while $\varGamma^{(d,d)}$ ($\varGamma_{m,n}$) denotes the Narain lattice for a $T^d$ torus (for the shifted circle). This has a nice interpretation as a Scherk-Schwarz partial supersymmetry breaking after one doubles the radius of the deformed $S^1$, so that the torus amplitude becomes
\eqn\SStorus{
\eqalign{
{\scr T} =& \int_{\sscr F} {d^2 \tau \over \tau_2^3} \, {1\over |\eta|^8}\, \left[
 |Q_o + Q_v|^2 \,\varGamma^{(4,4)}\, \left(  \varGamma_{m,2n} + \varGamma_{m+{1\over 2},2n} \right) \right.
\cr
 &+ |Q_o - Q_v |^2 \, \left| {2\eta\over \theta_2} \right|^4 \, \left(  \varGamma_{m,2n} - \varGamma_{m+{1\over 2},2n} \right) 
\cr
&  + 16\, |Q_s + Q_c |^2 \, \left| {\eta\over\theta_4}\right|^4\, 
\left( \varGamma_{m,2n+1} + \varGamma_{m+{1\over 2},2n+1 } \right)
\cr
& \left.+ 16 \, |Q_s - Q_c |^2 \, \left| {\eta\over\theta_3}\right|^4\, 
\left( \varGamma_{m,2n+1} - \varGamma_{m+{1\over 2},2n+1 }\right) \right] \varGamma^{(1,1)} \,.
\cr}
}
Standard supersymmetric orientifold projections of this interpolating type IIB configuration have already been studied in \adds, however we are interested now in non-BPS configurations with ${\rm O}p_+$ planes and for this reason, as in \bsb , we combine the world-sheet parity $\varOmega I_{45} \, (-1)^{F_{\rm L}}$, where $I_{45}$ denotes a simultaneous inversion along the $X_4$ and $X_5$ coordinates, with an automorphism $\sigma$ that reverts the contribution of the twisted sector. The Klein-bottle amplitude is then
\eqn\SSKlein{
{\scr K} = {\textstyle{1\over 4}} \int_0^\infty {d\tau_2 \over \tau_2^3} \, {1\over \eta^4}\, \left[ (Q_o + Q_v) (P^{(4)} + W^{(4)}) \, W_{2n} - 2\times 16 \, (Q_s + Q_c) {\eta^2 \over \theta_4^2}\, W_{2n+1} \right] W_n \,,
}
where, as usual, $P$ and $W$ denote the truncation of the Narain lattice to pure momenta and to pure winding zero modes. After an $S$ modular transformation to the tree-level channel, this amplitude clearly spells-out the geometry of O-planes: this interpolating orientifold contains two ${\rm O}7_-$ planes both with $X ^5 = 0$, together with 32 ${\rm O}3_+$ planes all at $X^5 = \pi R$, and dislocated at the 32 fixed points of the $T^4$ and of the spectator $S^1$. 

As expected, the open-string sector needed to cancel R-R tadpoles involves $N=16$ D7 and $M=16$ $\overline{{\rm D}3}$ branes, whose spectra are encoded in the annulus
\eqn\SSannulus{
{\scr A} = {\textstyle{1\over 2}} \int_0^\infty {dt \over t^3} \, {1\over \eta^4} \left[ 
\left( N^2\, P^{(4)} + M^2\, W^{(4)} \right) \, (Q_o + Q_v ) W_n
+ 2 NM\, (Q_s + Q_c) \, {\eta^2 \over \theta_4^2}\, W_{n+{1\over 2}} \right] W_n \,,
}
and M\"obius-strip
\eqn\SSMoebius{
\eqalign{
{\scr M} =&   - {\textstyle{1 \over 2}}  \int_0^{\infty} {d t \over t^3} \, {1\over \hat\eta^4} 
\Biggl[ N \, P^{(4)} \, (\hat V_4 \hat O_4 + \hat O_4 \hat V_4 - \hat S_4
\hat S_4 - \hat C_4 \hat C_4) \, W_{2n}
\cr
& -  M \, W^{(4)} \, ( \hat V_4 \hat O_4 + \hat O_4 \hat V_4 + \hat S_4
\hat S_4 + \hat C_4 \hat C_4)\,  W_{2n} 
\cr
& -   N \, (\hat V_4 \hat O_4 - \hat O_4 \hat V_4 + \hat S_4 \hat S_4 -\hat  C_4 \hat C_4) 
\left( {2 \hat \eta \over \hat \theta_2} \right)^2 \, W_{2n+1}
\cr
& +  M \, (\hat V_4 \hat O_4 - \hat O_4 \hat V_4 - \hat S_4 \hat S_4 + \hat C_4 \hat C_4)  \left( {2 \hat \eta \over \hat \theta_2} \right)^2 W_{2n+1} \Biggr] W_n
\cr}             
}
amplitudes. At the massless level the D7 branes comprise a full ${\scr N}=4$ vector supermultiplet in the adjoint of SO(16), while the $\overline{{\rm D}3}$ branes are non-supersymmetric and comprise vectors and six scalars in the adjoint of a USp(16) gauge group and four Weyl fermions in the reducible anti-symmetric representation ${\bf 120} = {\bf 119} + {\bf 1}$. The D7--$\overline{{\rm D}3}$ strings are here massive as a result of our choice of displacing the branes close to their homologous O-planes that in this model are geometrically separated.

Also in this case the configuration is unstable, although tachyon free, and is driven towards a strongly coupled regime. After the $\overline{{\rm D}3}$ are displaced in the bulk sufficiently far from the O-planes and from the D7 branes, one can use the same arguments based on S-duality and describe this model with $g_{\rm s} \gg 1$ in terms of a weakly coupled configuration where the 32 ${\rm O}3_+$ planes are traded for 32 $\widetilde{{\rm O}3}_-$ ones  $\sim 32 ({\rm O}3_- \ {\rm planes}+ \ {\rm stuck}\ D3$ branes. The sixteen bulk $\overline{{\rm D}3}$ branes can annihilate half of the stuck D3 ones and yield a fully supersymmetric configuration. The resulting massless spectrum has ${\scr N} =4$ supersymmetry and gauge group ${\rm SO} (16) \times {\rm SO} (16)$ as in the model in \adds, that was argued to be related to the heterotic M-theory of Horava and Witten \hw.  

It would be interesting to gain also some understanding of the strongly coupled regime of more general models with brane supersymmetry breaking, where the closed-string sector and presumably the final weakly coupled D-brane configuration have reduced supersymmetry. However, our arguments are based on the ${\rm SL} (2,\Bbb{Z})$ duality of type IIB, that is well established for ${\scr N}=4$ theories but not fully understood for non-maximally supersymmetric models. Although in principle it is not applicable to non-supersymmetric environments, in the models we have analysed in this letter S duality is only marginally broken since, if the antibranes are placed in the bulk, the configurations preserve to leading order sixteen supercharges, so that the strongly coupled regime is partly under control.

\vskip 36pt

\noindent
{\bf Acknowledgment.} We would like to thank Costas Bachas, Gianfranco Pradisi and Angel Uranga for discussions. 
C.A. thanks CPhT-Ecole Polytechnique and E.D thanks Univ. of Warsaw  for
hospitality during the completion of this work. Work partially supported by the CNRS PICS \#~2530 and 3059, RTN contracts MRTN-CT-2004-005104 and MRTN-CT-2004-503369, the European Union Excellence Grant, MEXT-CT-2003-509661 and the EC contract                                    MTKD-CT-2005-029466.

\listrefs

\end